# Data Driven Automatic Electrical Machine Preliminary Design with Artificial Intelligence Expert Guidance

Yiwei Wang, Tao Yang, IEEE senior member, Hailin Huang, Tianjie Zou, Jincai Li, Nuo Chen, Zhuoran Zhang, IEEE senior member

*Abstract*—This paper presents a data-driven electrical machine design (EMD) framework using wound-rotor synchronous generator (WRSG) as a design example. Unlike traditional preliminary EMD processes that heavily rely on expertise, this framework leverages an artificial-intelligence based expert database, to provide preliminary designs directly from user specifications. Initial data is generated using 2D finite element (FE) machine models by sweeping fundamental design variables including machine length and diameter, enabling scalable machine geometry with machine performance for each design is recorded. This data trains a Metamodel of Optimal Prognosis (MOP)-based surrogate model, which maps design variables to key performance indicators (KPIs). Once trained, guided by metaheuristic algorithms, the surrogate model can generate thousands of geometric scalable designs, covering a wide power range, forming an AI expert database to guide future preliminary design. The framework is validated with a 30kVA WRSG design case. A prebuilt WRSG database, covering power from 10 to 60kVA, is validated by FE simulation. Design No.1138 is selected from database and compared with conventional design. Results show No.1138 achieves a higher power density of 2.21 kVA/kg in just 5 seconds, compared to 2.02 kVA/kg obtained using traditional method, which take several days. The developed AI expert database also serves as a high-quality data source for further developing AI models for automatic electrical machine design.

## I. Introduction

Climate change is driving the shift to net-zero emission technologies, replacing conventional combustion engines with electric propulsion systems [1][2]. This transition increases the demand for both the quantity and performance of electrical machines, accelerating design cycles and the need for high-quality machine design schemes [3][4]. As Industry 4.0 evolves, AI applications have advanced data-driven design methodologies, meeting the growing demand for efficient electrical machine design (EMD) [5].

Despite these advancements, the typical EMD process remains largely equations-based and experience-driven, consisting of three stages: design specification clarification (S), preliminary design solution identification (P), and final detailed design solution (F) [6], as illustrated in Figure 1. The entire design process is referred to as S2PF.

The S2PF process has evolved through two phases: manual design lacking automation (LA) and semi-automation (SA), and now progressing towards full automation (FA) (Fig. 1). Initially, in the LA phase, design relied on designer experience, simplified formula, and repetitive experiments, making it time-consuming and inaccurate [7]. With the advent of design software like Maxwell and the integration of AI algorithms, the process transitioned into SA phase. During this phase, In the S2P (specification to preliminary design), software uses analytical methods or FE techniques to estimate the machine performance [8]. In the P2F, optimization tools are employed to identify the optimal design, involving surrogate algorithms and metaheuristic algorithms[9]. The surrogate algorithm is used to estimate performances, e.g. output voltages, more rapidly than FE model. Metaheuristic algorithms can greatly reduce the iteration cycles to derive an optimized solution rather than entirely relying on experience. Commonly surrogate algorithms include polynomial response surface [10], backpropagation neural network [11] and radial basis function [12] etc. Metaheuristic algorithms applied in EMD include evolutionary algorithm [13], particle swarm algorithm [14] and ant colony algorithms [15].

Recent studies have adopted optimization using surrogate models together with metaheuristic algorithms. In [16], the Kriging-aided particle swarm algorithm is used to optimise an 85kW permanent magnet motor, focusing on magnet usage and torque. In [17], a response surface surrogate model relates cogging torque to stator parameters, and the cuckoo search algorithm is applied to reduce cogging torque. In contrast, [18] implements optimisation using FE models to achieve objectives but with longer computation times compared to surrogate models [17]. Thus, combining surrogate algorithms with metaheuristic algorithms reduces calculation time and designer involvement, automating the P2F process.

Yiwei Wang, Tao Yang, Hailin Huang, Tianjie Zou are with the Power Electronics, Machines and Control (PEMC) Institute, University of Nottingham, Nottingham NG7 2GT, U.K.(e-mail: yiwei.wang1@nottingham.ac.uk; tao.yang@nottingham.ac.uk;hailin.huang@nottingham.ac.uk;tianjie.zou@nottingham.ac.uk).

Jincai Li is with the College of Automation Engineering, Nanjing University of Aeronautics and Astronautics, Nanjing, 210016, China (e-mail: cae_vsvf@nuaa.edu.cn).

Nuo Chen, is with the Institute of Electrical Engineering, Karlsruhe Institute of Technology, 76131 Karlsruhe, Germany. (e-mail: nuo.chen@kit.edu)

Zhuoran Zhang is with the College of Automation Engineering, Nanjing University of Aeronautics and Astronautics, Nanjing, 210016, China (e-mail: apsc-zzr@nuaa.edu.cn).

Despite progress in automating the P2F process, the S2P still heavily relies on expert experience and trial-and-error FEA calculations. To transition from semi-automation to full automation, automating the S2P is crucial, as shown in the bottom green layer in Fig. 1.

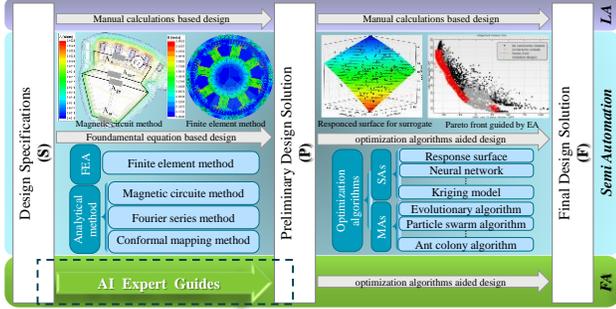

Fig. 1. The automation trend of S2PF.

To achieve this, a data-driven EMD method called AI Expert Guides is proposed. Following proposed design scaling rules, hundreds of scaled machine FE models are established and simulated to generate training data. This data is then used to train an AI expert surrogate model. Using the surrogate model, numerous design solutions are generated through parameter sweeping and clustered to form an AI expert database. This database enables immediate retrieval of preliminary design solutions based on given specifications, eliminating days of trial and error.

The main contributions of this paper are:

1. Proposing a data-driven AI expert to automatically guide preliminary machine design.

2. Establishing a framework for end-to-end mapping between machine geometry parameters (preliminary design) and design specifications (power, mass, efficiency, etc.) using surrogate models.

3. Introducing baselines, boundaries, and correlation functions as scaling rules to constrain geometry scaling during sample generation.

4. Developing an AI expert database for WRSG to support future preliminary design and AI model training study.

The paper is organized as follows: Section II introduces the fundamental principles of proposed automatic EMD method, including the mapping between training data inputs and outputs and the framework for developing MOP-based AI expert guides, along with a further introduction to the MOP algorithm. Section III details the method's development process: data generation, surrogate model training, design database generation and AI-guided design realization. Section IV presents the case of automatic electromagnetic design of a wound rotor synchronous generator (WRSG) using the proposed method. Section V verifies the AI-guided WRSG design solution using FEA and compares it to the original WRSG design solution used in the prototype, with its generating performance tested for FEA modelling verification.

## II. FUNDAMENTAL PRINCIPLES OF PROPOSED AUTOMATIC ELECTRICAL MACHINE DESIGN METHOD

### A. Mapping between machine specification requirement and machine design parameters

The performance of an electric machine from one specific design can be expressed as:

$$S = F(X, V) \quad (1)$$

where $X = (x_1, x_2, ..., x_n)$ is a set of geometry parameters, $V = (v_1, v_2...v_3)$ is a set of electromagnetic parameters, $S = (s_1, s_2, ..., s_h)$ is a set of performance specifications, $F$ is a non-linear function which maps the parameters $X$ and $V$ to its performance specifications $S$. Hence, for a specific set of parameters $X$ and $V$, the machine performance $S$ is uniquely determined. However, the EMD is an inverse process, i.e. to find geometry parameters $X$ for a specific performance requirement $S$. For a specific machine specification requirement $S$, the geometry parameter $X$ in (1) can be resolved with [19]:

$$X = F^{-1}(S, V) \quad (2)$$

It is noted that, for one specific $S$ (such as power, torque, weight, speed requirements etc) and $V$ (such as rated current density, rated magnetic density), there could be multiple sets of designs, i.e. different sets of $X = (x_1, x_2, ...x_n)$ including outer diameter, inner diameter, length etc, can fulfil. In other words, multiple machine designs $X$ exist for one specification requirement $S$ with the same electromagnetic load $V$. For one specific kind of machine design process, e.g. aircraft starter/generator, the electromagnetic load is kept the same while geometric dimensions are adjusted to achieve specifications, ensuring maximum material utilization. Hence, the mapping relation between specifications and parameters is from $X$ to $S$, which direction will be followed as training logic, $X$ used as input and $S$ used as output. The relation between $X$ and $S$ can be shown in Fig. 2.

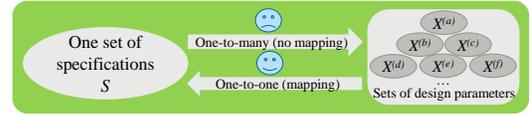

Fig. 2. The relation between specifications and the fundamental geometry design parameters

### B. Framework of automatic electrical machine design

The automatic EMD process fundamentally relies on development of data-driven design methodology. The data generation process is based on the fact that, for a given geometry parameter $X = (x_1, x_2, ...x_n)$ and specific electromagnetic load $V$, there is only one set of performance specification $S = (s_1, s_2, ...s_n)$ mapped. Additionally, concepts such as boundaries, baseline, and correlation function are proposed as scaling rules to restrict data generation, ensuring data obtained is suitable for use as training samples. Thus, through scaling the machine geometry under pre-defined rules, and evaluating the corresponding performance for each scaled machine design by FEA simulation, massive data can be generated.

With this generated data, an AI surrogate model can be trained and developed to generate AI expert database. The database contains hundreds of design solutions, allowing for the direct retrieval of solutions that satisfy project specifications, eliminating the need for time-consuming trial-and-error in the preliminary design phase. The surrogate model is trained based on the Metamodel of Optimal Prognosis algorithm (MOP), which is explained in the following section.

### C. Metamodel of optimal prognosis algorithm

MOP algorithm is an automatic approach of variable reduction and searching for the most appropriate surrogate model to map input variables (machine geometry X in our case) to one specific output (one of performance $s_j$ from S in our case) [20].

During the training process, a significant filter is used to select optimal input variables for one specific output. The correlation coefficient is defined as the linear coefficient of correlation $\rho_{ij}$ between the fit of regression $\hat{s}(x_i)$ and the variable $x_i$ on the samples $x_i^{(k)} x_j^{(k)}$, as defined in (3)

$$\rho_{ij} = \frac{1}{N-1} \frac{\sum_{k=1}^{N}\left(\hat{s}^{(k)}(x_i)-\mu_{\hat{s}(x_i)}\right)\left(x_j^{(k)}-\mu_{x_j}\right)}{\sigma_{\hat{s}(x_i)}\sigma_{x_j}} \quad (3)$$

where N is the total sample amount, $\mu$ is the mean value, σ is the population variance.

Generally, if $\rho_{ij}$ is greater than 0.7, this means the performance $s_j$ in S is strongly corelated with $x_i$ in X. Whereas less than 0.3 means weaking correlation between $s_j$ and design parameter $x_i$.

To find the most appropriate metamodel, it is essential to evaluate the accuracy of the predictions and assess the quality of an approximation of each metamodel applied. This is typically through using an additional test data set. The agreement between the actual test data and the predictions generated by the metamodel is quantified using a metric known as the coefficient of prognosis (*CoP*):

$$CoP = \left(\frac{E|S_{Test} \cdot \hat{S}_{Test}|}{\sigma_{S_{Test}}\sigma_{\hat{S}_{Test}}}\right)^2$$

$$= \left(\frac{\sum_{k=1}^{N}(s^{(k)}-\mu_s)(\hat{s}^{(k)}-\mu_{\hat{s}})}{(N-1)\sigma_s\sigma_{\hat{s}}}\right)^2 \quad (4)$$

$$0 \leq CoP \leq 1$$

Finally, the metamodel with maximum *CoP* is chosen as the optimal metamodel for each approximated response quantity in the AI expert surrogate model training process.

III. AUTOMATIC ELECTRICAL MACHINE DESIGN METHOD

The development of our proposed automated machine preliminary design tool includes four steps: data generation and collection, surrogate model training, database generation, and guides realisation. Each stage will be explained in this section.

A. Stage 1: Data generation and collection

To facilitate data generation, baseline, boundary, and correlation functions are proposed as scaling rules to guide the finite element (FE) model building process for each scaled machine. Based on the defined baseline, parameters with significant impact on geometry such as diameters and length e.g. are selected as fundamental variables $X=(x_1, x_2, ...x_n)$, and all other parameters $M=(m_1, m_2, ..., m_m)$ are depending on X by correlation functions $g_m$:

$$m_m = g_m(X) \quad (5)$$

With the geometry parameter set [X, M] and topology defined by the baseline, the FE model is constructed to estimate performance $P = (p_1, p_2, ...p_z)$. Through hundreds of sweeps across the flexible design space of variable X, sample data are generated. Given these correlation functions, M varies proportionally with X; thus, as X changes, the geometry scales accordingly. This process results in hundreds of different scaled machine geometries [X, M] and their corresponding performances P. The data sets can be represented as: $X=\{X^{(1)}, X^{(2)}, ...X^{(N)}\}$, $M=\{M^{(1)}, M^{(2)}, ..M^{(N)}\}$, $P=\{P^{(1)}, P^{(2)}, ...P^{(N)}\}$, where N is the number of sweeps, each $X^{(i)}= (x^{(i)}_1, x^{(i)}_2, ..., x^{(i)}_n)$, $M^{(i)}= (m^{(i)}_1, m^{(i)}_2, ..., m^{(i)}_m)$, $P^{(i)}=(p^{(i)}_1, p^{(i)}_2, ...p^{(i)}_z)$. Note that to enabling a broad machine geometry scaling process, the variable design space is significantly larger than that used in optimization process. The complete process of data generation and collection is illustrated in Fig. 3..

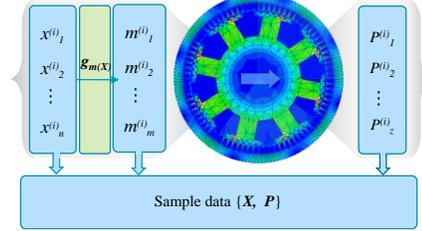

Fig. 4. Automated electric machine design tool development, stage 1: data generation and collection

Taking a wound-rotor synchronous machine (WRSM) as an example, the defined design boundaries, fundamental design parameters X, dependent parameters M and the performance index are shown in TABLE I.

TABLE I
DATA TYPES IN SAMPLE COLLECTING

| Boundaries B | Parameters | | Performance P |
|---|---|---|---|
| | Fundamental parameters X | Dependent parameters M | |
| $n_{max}$ | $D_1$ | $D_2$ | $P_{out}$ |
| $J_{max}$ | $N_a$ | $N_s$ | W |
| $A_{fmax}$ | PBH | PBW | T |
| … | … | … | … |
| $V_{rated}$ | $A_{ir}$ | $N_p$ | η |
| $f_{rated}$ | L | $N_f$ | THD |

From TABLE I, the design boundaries B include parameters such as maximum speed $n_{max}$, maximum armature current density $J_{max}$, maximum field winding current $A_{fmax}$, rated voltage $V_{rated}$, rated frequency $f_{rated}$. The variables X are fundamental variables of geometry parameters, such as the armature diameter $D_1$, the series turns per phase $N_a$, the pole body height *PBH*, the length of airgap $A_{ir}$, the length of laminations *L*. M is the dependent parameters, for example stator outer diameter $D_2$, number of slots $N_s$, pole body width *PBW*, number of poles $N_p$, turns of field windings per pole $N_f$. The variable P is the machine performance, such as output power $P_{out}$, weight *W*, shaft torque *T*, electromagnetic efficiency η, total harmonic distortion *THD*.

Through sweeping X within the design space, various scaled WRSM designs can be derived. These designs are then simulated using FE software to evaluate their performance index data. The resulting data sets can be divided into training and testing set for use in surrogate model training and evaluation process. This process will be further explained in the next section.

B. Stage 2: MOP based surrogate model training

With the data derived from step 1, a reliable surrogate model can be derived. Within this paper, the metamodel is used for surrogate model development. The developed metamodel uses three surrogate algorithms including polynomial least square, moving least squares, and isotropic Kriging. This metamodel is referred to as metamodel optimal prognosis (MOP). During the training process, MOP from different algorithms (PLS, MSL and Kriging) are used to map all the fundamental design variable sets $X = [X^{(1)}, X^{(2)}, …X^{(N)}]$ to one specific performance specification $p_i$. The surrogate model with maximum CoP is then used to map $X$ to $P_i$. Repeating this process through all the performance specifications, the surrogate model has thus been derived. The data training process is shown in Fig. 5.

Taking the wound-rotor synchronous machine as an example, the surrogate model is essentially mapping the independent variables $X = [D_1, N_a, … A_{ir}, L]$ and performance specification $P_{out}, W,…, \eta, THD$, respectively.

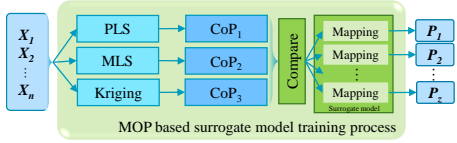

Fig. 5. Automated electric machine design tool development, stage 2: surrogate model training.

### C. Stage 3: AI expert database generation

With the trained AI surrogate model, numerous EMD solutions can be further generated to expand the knowledge base of AI expert through more detailed sweeping of design space. In Fig. 6, the database generation process begins by identifying the design variables $X$, from design space. Based on these variables, the corresponding performances $P$ and dependent parameters $M$ are calculated by surrogate model and correlation functions respectively. After $N$ times sweeping, the AI expert database which consists of $N$ solutions obtained. As these solutions are generated directly from the developed surrogate models and correlation functions, they can be derived in a much faster way.

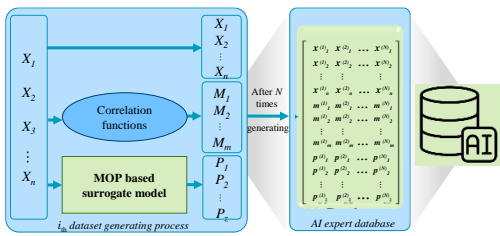

Fig. 6. Automated electric machine design tool development, stage 3: AI expert database generation.

### D. Stage 4: Design with AI Expert Guidance

With the massive EMD solutions generated in stage 3, for a specific design requirement, design solutions can be searched in the obtained database, as shown in Fig. 7. This process is named as the AI expert guides. Table II lists the given specifications of $P_{out}^{(i)}$, $W^{(i)}$, $\eta^{(i)}$, $D2^{(i)}$, and to achieve these specifications the solution $i$ including variables $X^{(i)}$ and parameters $M^{(i)}$ are searched from the AI expert database. Taking WRSG as an example, the parameters for solution $i$ are listed in Table III. It is noted that the specifications are provided by the EMD designer, and it composes not only performances but also parameters.

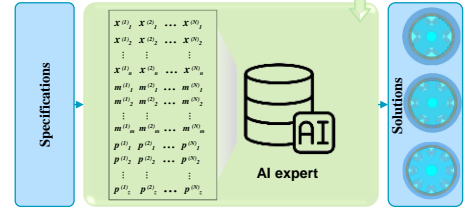

Fig. 7. Automated electric machine design tool development, stage 4: AI expert guides realization.

TABLE II
AI EXPERT GUIDES: SPECIFICATIONS FINDING SOLUTIONS

| Specifications($i$) | Solution($i$) |
|---|---|
| $P_{out}^{(i)}$ | |
| $W^{(i)}$ | $X^{(i)}$ |
| $\eta^{(i)}$ | $M^{(i)}$ |
| $D_2^{(i)}$ | |

TABLE III
DATA TYPES FOR SOLUTION I IN AI EXPERT

| Solution | $X^{(i)}$ | $M^{(i)}$ | $P^{(i)}$ |
|---|---|---|---|
| | $D_1^{(i)}$ | $D_2^{(i)}$ | $T^{(i)}$ |
| | $N_{a(i)}$ | $N_{s(i)}$ | |
| $i$ | … | … | |
| | $A_{ir}^{(i)}$ | $N_p^{(i)}$ | … |
| | $L^{(i)}$ | $N_f^{(i)}$ | $THD^{(i)}$ |

The entire AI based EMD framework is summarized in Fig. 8. It can be seen that the four stages are well integrated with each other starting from data generation to an automated machine design output. The entire process enables an automatic EMD concept. It should be noted that by applying a different baseline, this method can be adapted for any topology. The implementation of such a design framework will be demonstrated in the following section.

## IV. CASE STUDY: ELECTROMAGNETIC AUTOMATIC DESIGN FOR WOUND ROTOR SYNCHRONOUS GENERATOR

Wound rotor synchronous generator (WRSG) is one of the most widely used electrical machines for electric power generation [21]. Due to their high demand and well-established mature technology, WRSGs are ideal candidates for building an standard database for preliminary design. Given that the proposed methodology focuses on preliminary design, this case study will primarily address the core physics involved in electrical machine design, specifically the electromagnetic design, to verify the methodology. In this section, a case study of WRSG electromagnetic design is presented. The main objective is to demonstrate a preliminary design meeting the specification requirements that can be obtained from the developed AI expert database. The specifications include output power $P_{out}$, weight $W$, efficiency $\eta$, , which are typically provided for WRSG development in a project, are outlined in TABLE IV.

TABLE IV
SPECIFICATIONS REQUIREMENTS FOR CASE STUDY

| Specifications | Values |
|---|---|
| Apparent power ($P_{out}$) | >30kVA |
| Weight ($W$) | <17kg |
| Efficiency ($\eta$) | >92% |
| Out diameter ($D_2$) | <205mm |
| Speed ($n$) | 6000r/min |

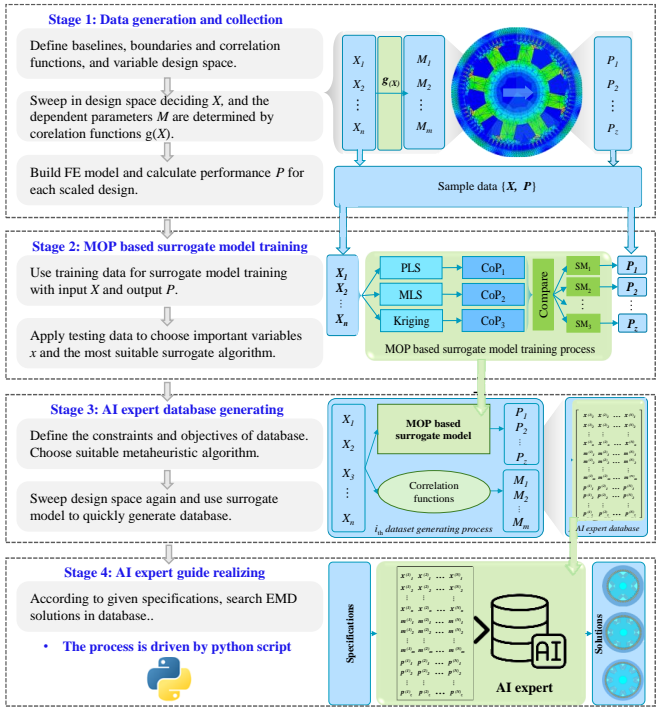

Fig. 8. The procedure of the proposed automatic EMD metho

## A. WRSG Scaling Data generation and collection

To establish the FE model of WRSG, a baseline model is referred to define material of stator and rotor lamination, rotor sleeves etc. (as we only focus on the geometry design at preliminary stage), as shown in Fig. 9(a). Furthermore, the baseline machine is a real machine that has been manufactured and tested, which allows to be used for comparison studies later on.

For a WRSG, six independent geometry parameters are identified as the fundamental variables including $D_1$ (armature diameter), $D_2$ (outer diameter), $L$ (length of laminations), $PBH$ (pole body height), $PBW$ (pole body width), $N_a$ (series turns per phase) as highlighted in Fig. 9(b). Other dependent parameters $M$, which depends on these 6 variables by correlation functions, such as the number of pole pairs $N_p$, and the number of stator slots $N_s$ etc, are listed in TABLE V. With these geometry parameters, a specific machine can be defined, and its corresponding FE model can be developed. The design boundaries $B$ such as maximum speed $n_{max}$, and maximum current density $A_{max}$ etc are defined as listed in TABLE V.

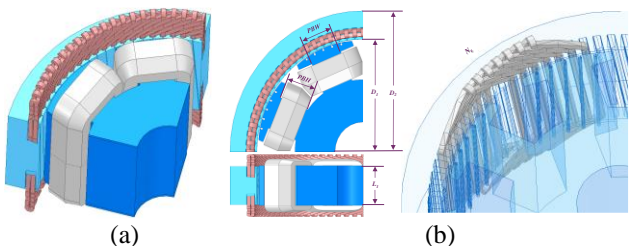

(a)          (b)
Fig. 9. 1/4 electromagnetic FEA model and variables of WRSG

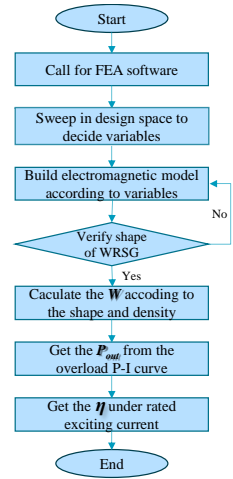

Fig. 10. The flow of Python script for WRSG performance calculation

The fundamental geometry variables are swept within their design space. For each sample point, an FE model of WRSG can be established and the performance of $P_{out}$, $W$, and $\eta$ can be derived through FE simulation. In order to generate data effectively, an automatic data generation process environment has been developed in Ansys OptisLang environment using python. As shown in Fig. 10, the developed scripts firstly call for the FE software (MotorCAD in our case). The machine design fundamental variables, dependent parameters, boundaries and correlation functions are defined as listed in TABLE V. With these parameters, the entire WRSG machine geometry is defined and will be validated to check if the design is achievable and valid. If the derived WRSG geometry is valid, the machine output power $P$, weight $W$ and efficiency $\eta$ will be calculated.

In this case study, a total of 400 different scaled geometry designs have been generated to simulate, which takes around 31 hours, and the simulated results are collected as sample data sets. After data generation and collection (Stage 1), a surrogate model can be trained (Stage 2). During the data training process, the design parameters $D_1$, $D_2$, $L$, $PBH$, $PBW$, $N_a$ are used as inputs of the surrogate model and the performance indices $P_{out}$, $W$, $\eta$ are defined as the outputs. This is further explained in the next section.

## B. Surrogate model development and analysis

During the surrogate model training process, the correlation of the design parameter (inputs) and the performance indices (outputs) need to be carefully observed and analysed. The correlation is defined as the CoP matrix as discussed before. The CoP of example design cases is shown in Fig. 11, which takes 3 mins to be trained from 400 samples. The CoP for each performance index (Weight, Power and Efficiency) correlated to all the inputs variables (L, PBH, PBW, D2, Na, $D_1$) is shown in the right column. As can be seen, the CoP values are all very high and over 93%. This means that the AI expert surrogate model can be with good accuracy to predict performance with given design data (L, PBH, PBW, D2, Na, D1). In Fig. 11, the correlation of each design parameter and performance index can also been identified. As can be seen, the weight $W$ is more relevant to the inner diameter of the machine ($D_1$) and the outer diameter of the machine ($D_2$) as well as Armature turns per phase per pole pair ($N_a$). For the output power $P$, the inner diameter $D_1$ has much more impact. For the efficiency $\eta$, the inner core length $L$ has the highest impact and machine inner diameter is also with relatively higher impacts compared with other four design parameters.

TABLE V
DESIGN BOUNDARIES, VARIABLES, AND DEPENDENT PARAMETERS FOR WRSG

| Parameter type | Symbols | Physical parameters | Units | Value/corelation function |
|---|---|---|---|---|
| $B_1$ | $V_{rated}$ | Rated voltage | V | 115 |
| $B_2$ | $n_{rated}$ | Rated speed | r/min | 6000 |
| $B_3$ | $f_{rated}$ | Rated frequency | Hz | 400 |
| $B_4$ | $A_{max}$ | Maximum armature current density | $Amm^{-2}$ | 16 |
| $B_5$ | $A_{fmax}$ | Maximum field current density | $Amm^{-2}$ | 12 |
| $B_6$ | $n_{max}$ | Maximum speed | r/min | 7200 |
| $X_1$ | $D_1$ | Inner diameter | mm | (100, 200) |
| $X_2$ | $D_2$ | Outer diameter | mm | (120, 250) |
| $X_3$ | $L$ | Length of iron core | mm | (40, 80) |
| $X_4$ | $PBH$ | Pole body height | mm | (20, 40) |
| $X_5$ | $PBW$ | Pole body weight | mm | (20, 40) |
| $X_6$ | $N_a$ | Armature turns per phase per pole pair | - | (5, 6, 7) |
| $M_1$ | $N_p$ | Number of pole pairs | - | $2*30* f_{rated} / n_{rated}$ |
| $M_2$ | $N_s$ | Number of stator slots | - | $N_a *3* N_p /2$ |
| $M_3$ | $N_f$ | Field turns per pole | - | $Drc/0.6$ |
| $M_4$ | $Ws$ | Slot width | mm | $N_a /( N_a +1)*2.2/3.8*((D_1+1)*pi/ N_s)$ |
| $M_5$ | $Drc$ | Rotor coil depth | mm | $PBH-0.1$ |
| $M_6$ | $Ds$ | Slot depth | mm | $0.25*(D_2-D_1)/2$ |
| $M_7$ | $PTW$ | Pole tip width | mm | $0.30*PBH$ |
| $M_8$ | $PTD$ | Pole tip depth | mm | $0.10*PBW$ |
| $M_9$ | $PSR$ | Pole surface radius | mm | $(D_1-1.4)/1*0.8$ |
| $M_{10}$ | $PSO$ | Pole surface offset | mm | $(D_1-1.0)/2*0.57$ |
| $M_{11}$ | $Dsh$ | Shaft diameter | mm | $D_1/2.075$ |
| $M_{12}$ | $Wac$ | Copper width | mm | Slot width-2*0.85 |
| $M_{13}$ | $Hac$ | Copper height | mm | (Slot depth-1.6)/2 |

## C. AI expert database generation

With the trained surrogate model, extensive data can be generated by sweeping design variables ($L, PBH, PBW, D_2, Na, D_1$) within their design space. In our example, we use the surrogate model to generate 9900 designs, and this can be achieved in 10 minutes.

With the design database derived from the extensive simulation of the surrogate model, these solutions can be projected into power-weight plane and a Pareto Front can be identified as shown in Fig. 12. There, 9,900 examples have been generated. The black points are the designs that satisfied efficiency constraint, i.e. $\eta > 92\%$, the grey points are the designs which violate this constraint, i.e. $\eta < 92\%$. The red points in Fig. 12 are the predicted Pareto Front of weight and power outputs. Designs at the Pareto Front will consist of the database, including design solutions with highest power density at a specific power rating requirement.

To validate the accuracy of the generated AI expert database, some red points recalculated by FE and the results are shown as green points in Fig. 12. It can be seen the pareto front of the green points overlap with the predict pareto front of the red points. A few designs from the expert library (i.e. designs at the Pareto front in Fig. 12, No.1138, No.3048, No.4625, and No.7049) have been selected. The resulting efficiency, weight and output power from FEA models are compared with those from surrogate models as shown in Fig. 13. It can be seen that results from these two models, i.e. Weight_OPT, Output_Power_OPT and Efficiency_OPT, are matched very well.

## D. AI expert guides implementation

From the derived AI expert database, we need to identify preliminary designs which meet the performance requirement as specified in TABLE IV, i.e. power ($P_{out}$) > 30kVA, weight ($W$) < 17kg and efficiency $\eta$ >92%. Six preliminary designs listed in TABLE VI are directly searched from database within 5s, which is much faster than several days' trial and error depending on experience. The details including variables and performances of 6 designs are given in Table VI, and based on that, the power density is calculated, which is one of most important requirements for aircraft electrical machine. Among these six design solutions, the solution of No.1138 is chosen as the final solution, as it has the highest power density 2.12kVA/kg, higher than the 2.00kVA/kg of the original FEA calculated solution(baseline).

TABLE VI
TYPICAL WRSG DESIGN SOLUTIONS GUIDED BY AI EXPERT

| Parameters | Type | Baselines | No.491 | No.548 | No.1023 | **No.1138** | No.4587 | No.5659 |
|---|---|---|---|---|---|---|---|---|
| $P_{out}$/kVA | P | 30.05 | 30.39 | 32.69 | 30.82 | **32.54** | 30.73 | 30.51 |
| W/kg | P | 15.11 | 14.89 | 16.65 | 15.04 | **15.36** | 16.62 | 16.63 |
| E/% | P | 94.45 | 94.21 | 93.99 | 93.96 | **94.03** | 88.72 | 94.32 |
| $D_2$/mm | X | 204.95 | 199.35 | 195.93 | 204.27 | **204.22** | 204.58 | 197.59 |
| L/mm | X | 70.04 | 72.38 | 76.12 | 70.27 | **71.49** | 71.00 | 74.13 |
| PBH/mm | X | 22.12 | 23.13 | 21.68 | 23.02 | **22.69** | 24.03 | 21.84 |
| PBW/mm | X | 22.36 | 24.13 | 23.96 | 24.74 | **25.03** | 26.72 | 23.75 |
| $D_1$/mm | X | 163.40 | 166.40 | 163.04 | 171.82 | **170.61** | 179.31 | 159.41 |
| $N_a$ | X | 7 | 7 | 6 | 7 | **7** | 6 | 6 |
| Power density/kVAkg$^{-1}$ | Obj. | 2.00 | 2.04 | 1.96 | 2.05 | **2.12** | 1.85 | 1.83 |

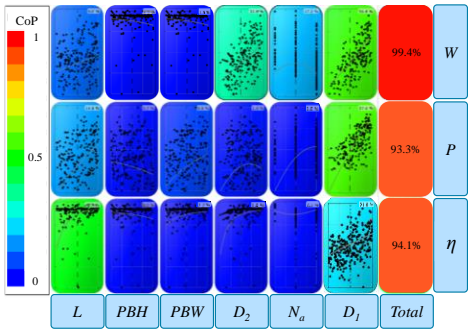

Fig. 11. CoP matrix of AI expert surrogate model

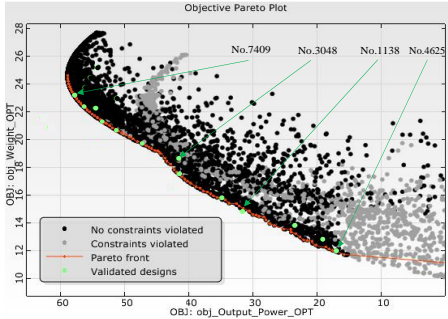

Fig. 12. Surrogate model generated design database with Pareto Front representing the optimised solutions which form an AI expert guided design database

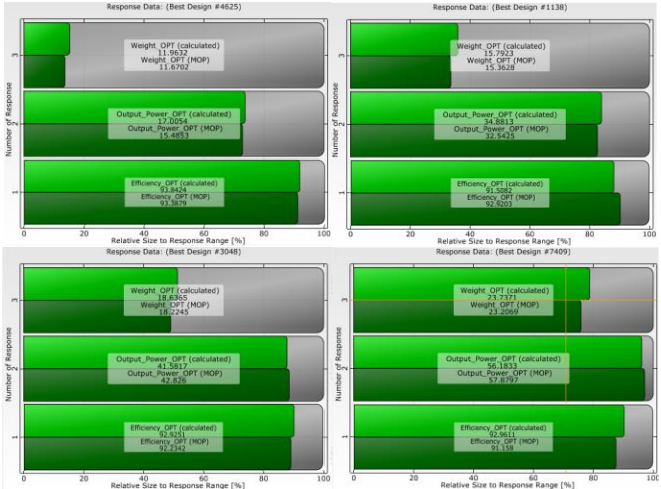

Fig. 13. Comparison between prediction and FEA calculation experimental validation

According to the design requirements in TABLE IV, an electrical machine has been designed and manufactured using conventional design methodology before our AI expert driven design has been proposed. The developed WRSG so far is shown in *Fig. 14.*

With the same requirements, a more effective and quicker design can be achieved using our AI expert with the design No.1138 been selected. The corresponding FE models of these two designs (one is the FE model of the existing design and the other is No.1138) have been compared and shown in TABLE VII. From TABLE VII, it can be seen that the power and weight from FEA and experimental results are very close in regard to the existing machine.

The data regarding the existing prototype is from real measured data with the WRSG machine mounted test bench in *Fig. 14.* The voltage and current wave of phase A from FEA model and measurement are shown in Fig. *15.* As can be seen, from the same operation point, the measured RMS of voltage is 117V and the RMS of current is 84.4A under a frequency of 400Hz, while the voltage is 115.8V and current 85.7A from its FE model. This gives us confidence that the FEA model can well represent the real hardware for WRSG performance estimation. On the other hand, performance of the FE model using No.1138 design data from AI expert is also shown in TABLE VIII. Comparing the two FE in the table, it can be seen that using our AI expert driven design can achieve a higher power rating (34.88kVA compared with 29.82kVA) and a higher power density (2.21kVA/kg compared with 2.00kVA/kg). This is really a promising result and gives us confidence to use our AI expert for future WRSG preliminary design and development.

TABLE VII
COMPARISON BETWEEN AI GUIDED AND ORIGINAL DESIGN

| Solutions | AI guided (No.1138) | | Existing design using conventional method | |
|---|---|---|---|---|
| | MOP | FEA | FEA | Experiment |
| $P_{out}$/kVA | 32.54 | 34.88 | 29.82 | 29.62 |
| W/kg | 15.36 | 15.79 | 15.11 | 15.21 |
| $\eta$ /% | 94.03 | 93.77 | 94.45 | - |
| $D_2$/mm | 204.22 | | | 204.95 |
| L/mm | 71.49 | | | 70.04 |
| PBH/mm | 22.69 | | | 22.12 |
| PBW/mm | 25.03 | | | 22.36 |
| $D_1$/mm | 170.61 | | | 163.40 |
| $N_a$ | 7 | | | 7 |
| Power density/kVAkg$^{-1}$ | 2.12 | 2.21 | 2.00 | 2.02 |

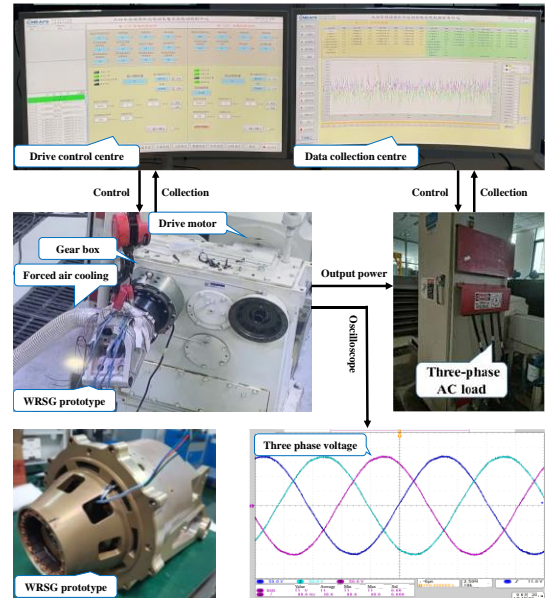

Fig.14. The established generating performance Test bench

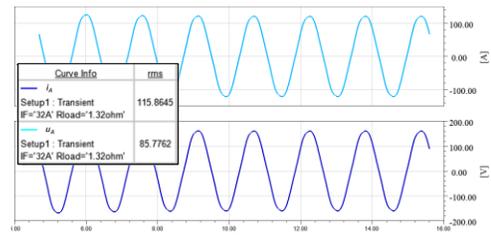

(A) The voltage and current wave of phase A simulated by FEA

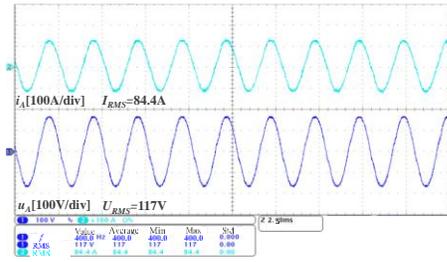

(B) The voltage and the current wave of phase A measured by oscilloscope

Fig.15. Phase voltage and current wave measured by an oscilloscope.

## V. Conclusion

With the aim of investigating the potential of data-driven based electrical machine design (EMD) concept, this paper proposed a framework of developing AI expert database to generate EMD preliminary solutions directly. For conventional EMD, it is a process that needs expertise and time-consuming FEA trial error. However, by harnessing the method of AI expert guides, the preliminary solutions can be directly obtained in seconds. Although the developing process takes around 31 hours, the result data can be used as standard database for future WRSG design. The case of 30kVA WRSG automatic design using AI expert guides has been studied. According to the given specifications, 6 design solutions were obtained from database. It costed 5s to search, which is faster than days-consuming of FEA based EMD. One of the 6 design solutions, No.1138, which has the highest power density, has been chosen and verified by FEA. It is shown that the error of output power and weight are 0.65% and 0.6%, respectively. Further, the AI expert guided design solution has a higher power density of 2.21kVA/kg compared to the original optimised WRSG design solution of 2.02kVA/kg, which is applied to the prototype and the experiment shows the performance has little error to FEA performance calculation. It is concluded that the proposed AI expert guided EMD method accelerates the process from specifications to preliminary design solutions and reduces the reliance from designer.